\documentclass[aps,prd,showpacs,twocolumn,showkeys,reprint]{revtex4-1}
\usepackage{amssymb}
\usepackage{amsmath}
\usepackage{amsfonts}
\usepackage{color}
\usepackage[usenames,dvipsnames]{xcolor}
\usepackage{float}
\usepackage{accents}
\usepackage{graphicx}
\usepackage{epstopdf}
\usepackage{bm}
\usepackage{times}
\usepackage{euscript}
\usepackage[colorlinks=true,pdfstartview=FitV,linkcolor=blue,citecolor=blue,urlcolor=blue,breaklinks=true]{hyperref}

\begin{document}
\title{Topological self-dual configurations in a Maxwell--Higgs model with a CPT-odd and Lorentz-violating nonminimal coupling}
\author{Rodolfo Casana}
\email{rodolfo.casana@gmail.com}
\author{Manoel M. Ferreira Jr.}
\email{manojr.ufma@gmail.com}
\author{Alexsandro Lucena Mota.}
\email{lucenalexster@gmail.com}
\affiliation{Departamento de F\'{\i}sica, Universidade Federal do Maranh\~{a}o, 65080-805,
S\~{a}o Lu\'{\i}s, Maranh\~{a}o, Brazil.}

\begin{abstract}
We have studied the existence of topological self-dual configurations in a
nonminimal CPT-odd and Lorentz-violating (LV) Maxwell--Higgs model, where the
LV interaction is introduced by modifying the minimal covariant derivative.
The Bogomol'nyi--Prasad--Sommerfield formalism has been implemented, revealing
that the scalar self-interaction implying self-dual equations contains a
derivative coupling. The CPT-odd self-dual equations describe
electrically neutral configurations with finite total energy proportional to
the total magnetic flux, which differ from the {charged} solutions of other
CPT-odd and LV models previously studied. In particular, we have investigated
the axially symmetrical self-dual vortex solutions altered by the LV
parameter. For large distances, the profiles possess general behavior similar
to the vortices of Abrikosov--Nielsen--Olesen. However, within {the} vortex
core, the profiles of the magnetic field and energy can differ substantially
from ones of the Maxwell--Higgs model depending if the LV parameter is negative
or positive.

\end{abstract}


\maketitle

\section{Introduction}

The study of magnetic vortices in condensed matter physics was established by
the seminal work {of} Abrikosov on superconductivity \cite{Abrikosov}, based {on}
the Ginzburg--Landau theory \cite{GLT}. The magnetic field in Type II
superconductors {forms} a bidimensional periodic structure known as Abrikosov's
vortex because {of} its great similarity with the Onsager and Feynman vortices
appearing in superfluid helium II \cite{Onsager}. In the field theory context,
magnetic vortex solutions were studied initially in the Maxwell--Higgs
electrodynamics by Nielsen and Olesen \cite{NOlesen} and by Schaposnik and de
Vega \cite{Vega}. The Abelian Higgs model, a relativistic generalization of
the Ginzburg--Landau theory of superconductivity, provides vortex solutions
endowed with quantized magnetic flux. From the {1980s}, the topological
Chern--Simons (CS) term has played an important role in gauge field theories
and condensed matter physics, with its effects being investigated both at the
classical and {the} quantum level. This way, the inclusion of the Chern--Simons term
in models, where the gauge field is minimally coupled to fermionic fields
\cite{Hyun} or to Higgs fields \cite{SV,HKJ}, allows to describe magnetic and
electrically charged vortices. Further, in many cases these vortices are
self-dual configurations saturating the Bogomol'nyi--Prasad--Sommerfield (BPS)
bound \cite{BPS}.

Since 1998 Lorentz symmetry breaking investigations have been considered
mainly in the framework of the standard model extension (SME) \cite{SME1}. The
SME is an extension of the usual standard model including new terms
or couplings between the LV backgrounds and the standard model fields. The LV
backgrounds arise as vacuum expectation values of tensor fields due to a
spontaneous Lorentz-symmetry breaking occurring in a theory at very high
energy scale \cite{SME1,SME2}. The study of the Lorentz-violating effects on the formation of topological defects was firstly considered in the solitonic solutions generated by scalar fields in Refs.
\cite{Bazeia1,Dutra1,Gomes1,Dutra2}. Topological defects arising in scenarios with spontaneous Lorentz-symmetry breaking triggered by
tensor fields were reported in Ref. \cite{Seifert}. The preliminary study
about vortex solutions in a Maxwell--Higgs electrodynamics with the
Carroll--Field--Jackiw CPT-odd term was reported in Ref. \cite{Baeta} and the
verification of the existence of BPS and charged vortices was shown in Ref.
\cite{Lazar}. The existence of BPS vortices in a Maxwell--Higgs model in the presence of CPT-even and LV gauge terms {{was} performed in
Refs. \cite{Miller1,Miller2}, being shown that LV coefficients
provide some characteristics which are not shared by the usual Maxwell--Higgs
vortices. Other studies about vortex configurations were developed
in the Maxwell--Higgs CPT-even LV model \cite{Hott, Belich,Sourrou1} and in the
context of the CPT-odd LV term \cite{Sourrou2,Alyson}. Recently, new effects
of LV terms on different kind of defects have been also investigated in the
context of oscillons and breathers generated in systems of coupled scalar
fields \cite{Correa1}, traveling solitons in Lorentz and CPT-odd models
\cite{Correa2}, and long-living, time-dependent and spatially localized field
oscillons configurations \cite{Correa3}.

Vortex configurations were also investigated in the context of the
Maxwell--Chern--Simons theories with nonminimal interactions. One of the first
studies \cite{Torres} with a nonminimal coupling was introduced by means of
the derivative,
\begin{equation}
D_{\mu}\phi=\left(  \partial_{\mu}-ieA_{\mu}-i(g/4)\epsilon_{\mu\nu\rho}%
F^{\nu\rho}\right)  \phi\mathbf{,}\label{NM0}%
\end{equation}
which implied BPS and nontopological solutions. An altered version of
this model, with a dielectric function $G(\left\vert \phi\right\vert )$
inserted in the derivative (\ref{NM0}), was also considered \cite{Ghosh},
besides other investigations in nonminimal models involving
other distinct aspects \cite{LM}.

It is well known that the inclusion of CPT-odd and/or CPT-even
Lorentz-violating terms in a determined field model can provide new features,
altering its properties. One way to include Lorentz violation is modifying the
kinetic sector of the fields. Another way is to introduce the Lorentz
violation factor via nonminimal couplings, that is, the new
interactions involving the fixed LV backgrounds and the
fields. Up {to} the moment, however, there is no investigation about
self-dual configurations in field theories endowed with nonminimal couplings
containing a 4-vector yielding a preferred direction in spacetime. Our aim is
to investigate such a possibility in a Maxwell--Higgs model modified by a
CPT-odd nonminimal covariant derivative (introduced in Refs.
\cite{petrov1,petrov2}), which includes a Lorentz-violating vector background.
The manuscript is organized as follows: In Sec. II, the theoretical framework
is established and its relevant equations are presented. In Sec. III, we
implement the BPS formalism with the aim at finding self-dual or BPS equations
describing electrically neutral topological configurations possessing finite
energy proportional to the magnetic flux. By using the axially symmetric
vortex \textit{Ansatz} it is shown the BPS vortices behave like the
Abrikosov--Nielsen--Olesen ones. In Sec. IV, we perform the numerical solution
of the self-dual equations and we do a detailed analysis of the solutions by
comparing the effects of the Lorentz-violation on the Maxwell--Higgs self-dual
solution. In Sec. V, we finalize with our remarks and conclusions.

\section{A CPT-odd and Lorentz-violating nonminimal Maxwell\color[rgb]{1,0,0}{--}Higgs model}

Lorentz-violating nonminimal couplings have been examined in an extended version of the minimal SME embracing higher order {derivative} terms in the
photon sector \cite{KMNM1} and in the fermion sector \cite{KMNM2}. Other types
of Lorentz-violating nonminimal couplings, representing new interactions
between photons and fermions and not contained in these latter nonminimal
extensions, were proposed as well. A CPT-odd nonminimal coupling of this kind
was first considered in Ref. \cite{NM1} in the context of the Dirac equation,
by means of the following extended covariant derivative:
\begin{equation}
\mathcal{D}_{\mu}=\partial_{\mu}-ieA_{\mu}+i\frac{g}{2}\epsilon_{\mu\nu
\alpha\beta}V^{\nu}F^{\alpha\beta},\label{NM1}%
\end{equation}
where $A_{\mu}$ is an Abelian gauge field and $F_{\mu\nu}=\partial_{\mu}%
A_{\nu}-\partial_{\nu}A_{\mu}$ is the respective strength tensor. Here, the
four-vector $V^{\mu}$ is a fixed background which breaks both the CPT and
Lorentz symmetry. The implications of this nonminimal coupling on fermionic
fields have been intensively examined in several aspects, including the
fermion--fermion ultrarelativistic scattering \cite{Charneski}, generation of
radiative corrections \cite{Petrov}, the induction of several types of
topological and geometrical phases \cite{Bakke}, and the dynamics of the
Aharonov--Casher--Bohm problem \cite{NMACB}.

Recently, it was proposed another CPT-odd and Lorentz-violating
nonminimal covariant derivative \cite{petrov1,petrov2},
\begin{equation}
\mathcal{D}_{\mu}=\partial_{\mu}-ieA_{\mu}+iF_{\mu\nu}\xi^{\nu},\label{NM2}%
\end{equation}
where $\xi^{\mu}$ is the fixed four-vector background responsible for
violating the CPT and Lorentz symmetries. The nonminimal coupling (\ref{NM2})
has been considered to analyze some aspects of the physics of light pseudoscalars or axionlike particles \cite{petrov1} and to generate higher-derivative LV contributions to the photonic effective action in a Quantum Electrodynamics scenario \cite{petrov2}.

Up to now, all investigations about solitonic configurations in
Lorentz-violating field models have been made by means of the modification of
the kinetic terms of the fields or by dimensional reduction. On other
side, our purpose is to analyze the effects of a CPT-odd and LV nonminimal
coupling in the self-dual configurations of (1+2)-dimensional Abelian Higgs
models. In this context, the LV nonminimal covariant derivative (\ref{NM1}),
when projected in planar configurations, becomes equivalent to the
Lorentz-invariant one given in Eq. (\ref{NM0}), whose solitonic configurations
have been already investigated in the literature. In order to turn this model
interesting, one could add CPT-even and LV terms to the Higgs sector, which is
not our aim now. On the other hand, the nonminimal coupling (\ref{NM2}) yields
a {distinct} scenario whose features were not studied yet, being this the reason
to be addressed here.

The CPT-odd and Lorentz-violating $(1+3)$-dimensional model, in which
our investigation {is} based, is defined by the following Lagrangian density:
\begin{equation}
\mathcal{L}=-\frac{1}{4}F_{\mu\nu}F^{\mu\nu}+\left\vert \mathcal{D}_{\mu}%
\phi\right\vert ^{2}-{U\left(  \left\vert \phi\right\vert ,\partial_{\mu
}\left\vert \phi\right\vert \right)  },\label{aa1}%
\end{equation}
where
\begin{equation}
\mathcal{D}_{\mu}\phi=\partial_{\mu}\phi-ieA_{\mu}\phi+iF_{\mu\nu}\xi^{\nu
}\phi,
\end{equation}
is the nonminimal covariant derivative of the Abelian Higgs field, $\xi^{\nu}$
is a CPT-odd Lorentz-violating vector and ${{U\left(  \left\vert
\phi\right\vert ,\partial_{\mu}\left\vert \phi\right\vert \right)  }}$ is an appropriate positive-definite interaction to be determined.

The gauge field equation of motion is%
\begin{equation}
\partial_{\nu}F^{\nu\mu}+\xi^{\nu}\partial_{\nu}\mathcal{J}^{\mu}%
=e\mathcal{J}^{\mu},\label{ggff}%
\end{equation}
whose current density,%
\begin{equation}
\mathcal{J}^{\mu}=i\left(  \phi\partial^{\mu}\phi^{\ast}-\phi^{\ast}%
\partial^{\mu}\phi\right)  -2eA^{\mu}\left\vert \phi\right\vert ^{2}%
+2F^{\mu\rho}\xi_{\rho}\left\vert \phi\right\vert ^{2},\nonumber
\end{equation}
is conserved, that is, $\partial_{\mu}J^{\mu}=0$.

The Higgs field equation is given by%
\begin{equation}
\mathcal{D}_{\mu}\mathcal{D}^{\mu}\phi-\partial_{\mu}\frac{\partial
U}{\partial\partial_{\mu}\phi^{\ast}}+\frac{\partial U}{\partial\phi^{\ast}%
}=0.
\end{equation}

\section{Planar and topological self-dual configurations}

In order to study the planar configuration of the Lorentz-violating
model (\ref{aa1}), we adopt the following projection: $\partial_{3}\phi=0$,
$A_{3}=0$, $\partial_{3}A_{\mu}=0$. The Greek indexes denote $\mu,\nu=0,1,2,$
while the Latin indexes, $j,k=1,2$.

In the static regime, Eq. (\ref{ggff}) provides the following planar Gauss law:%
\begin{equation}
\partial_{k}\partial_{k}A_{0}+\xi_{k}\partial_{k}\mathcal{J}_{0}%
=-e\mathcal{J}_{0},\label{gauss}%
\end{equation}
with $J_{0}$ given by%
\begin{equation}
\mathcal{J}_{0}=-2eA_{0}\left\vert \phi\right\vert ^{2}+2\xi_{j}\left(
\partial_{j}A_{0}\right)  \left\vert \phi\right\vert ^{2}.
\end{equation}
Similarly, the planar Ampere law reads%
\begin{equation}
\epsilon_{kj}\partial_{j}B-\xi_{j}\partial_{j}\mathcal{J}_{k}=e\mathcal{J}%
_{k},\label{ampere0}%
\end{equation}
where $\epsilon_{kj}$ is the two-dimensional Levi--Civita
symbol ($\epsilon_{12}=1=-\epsilon_{21}$). Here, the planar magnetic
field is defined by $B=F_{12}$, and $J_{k}$ reads as%
\begin{align}
\mathcal{J}_{k}  & =i(\phi\partial_{k}\phi^{\ast}-\phi^{\ast}\partial_{k}%
\phi)-2eA_{k}|\phi|^{2}\label{jjk}\\[0.15cm]
& +\xi_{0}\left\vert \phi\right\vert ^{2}\partial_{k}A_{0}-2\epsilon_{kj}%
\xi_{j}B|\phi|^{2}.\nonumber
\end{align}

It is clear from the Gauss law that $eJ_{0}$ stands for the
electric charge density, so that the total electric charge of the configurations is%
\begin{equation}
Q=e\int d^{2}x\,\mathcal{J}_{0}.
\end{equation}
which is shown to be null ($Q=0$) by integration of the Gauss
law under suitable boundary conditions for the fields at infinity, i.e., $A_{0}\rightarrow0$ and $\phi\rightarrow cte$. Therefore, the field configurations will be electrically neutral, like it happens in the usual Maxwell--Higgs model.

The fact the configurations are electrically neutral is compatible with the gauge condition, $A_{0}=0$, which satisfies identically the Gauss law (\ref{gauss}). With the choice $A_{0}=0$, the static and electrically neutral configurations are described by two equations. The first one is the planar Ampere law
\begin{equation}
\epsilon_{kj}\partial_{j}B-\xi_{j}\partial_{j}J_{k}=eJ_{k},\label{ampere}%
\end{equation}
where $J_{k}$ obtained from (\ref{jjk}) is
\begin{equation}
J_{k}=i(\phi\partial_{k}\phi^{\ast}-\phi^{\ast}\partial_{k}\phi)-2eA_{k}%
|\phi|^{2}-2\epsilon_{kj}\xi_{j}B|\phi|^{2},
\end{equation}
and we observe that the dependence in the LV parameter $\xi_{0}$ has disappeared.

The second equation is the Higgs field one, which becomes
\begin{equation}
\mathcal{D}_{k}\mathcal{D}_{k}\phi+\partial_{k}\frac{\partial U}{\partial\partial_{k}\phi^{\ast}}-\frac{\partial U}{\partial\phi^{\ast}%
}=0,\label{higgs}%
\end{equation}
where the nonminimal covariant derivative $D_{k}\phi$ is written as%
\begin{equation}
\mathcal{D}_{k}\phi=\partial_{k}\phi-ieA_{k}\phi-i\epsilon_{kj}B\xi_{j}%
\phi.\label{DK}%
\end{equation}

By carefully observing Eqs. (\ref{ampere}) and (\ref{higgs}), we clearly note that the LV components $\xi_{0}$, $\xi_{3}$ do not participate in the formation of the electrically neutral configurations.

In order to implement the BPS formalism, we write the energy for the static and electrically neutral configurations, $E=\int d^{2}x~(-\mathcal{L})$, by considering the gauge $A_{0}=0$, that is,
\begin{equation}
E=\int d^{2}x\left(  \frac{1}{2}B^{2}+\left\vert \mathcal{D}_{k}%
\phi\right\vert ^{2}+U\right)  .\label{eenn}%
\end{equation}

After some algebraic manipulations, it is possible to establish the following identity:%
\begin{equation}
\left\vert \mathcal{D}_{k}\phi\right\vert ^{2}=\left\vert \mathcal{D}_{\pm
}\phi\right\vert ^{2}\pm eB\left\vert \phi\right\vert ^{2}\pm B\xi_{j}%
\partial_{j}\left\vert \phi\right\vert ^{2}\pm\partial_{i}\mathbb{J}_{i},
\end{equation}
where we have defined
\begin{align}
\mathcal{D}_{\pm}\phi &  =\mathcal{D}_{1}\phi\pm i\mathcal{D}_{2}%
\phi,\\[0.2cm]
\mathbb{J}_{i} &  =\frac{1}{2}\epsilon_{ik}J_{k}-\xi_{i}B\left\vert
\phi\right\vert ^{2},
\end{align}
with $\mathcal{D}_{k}\phi$\ given by Eq. (\ref{DK}). Such an identity allows to obtain the expression for the energy,
\begin{align}
E &  =\int d^{2}x\left[  \frac{1}{2}\left(  B\mp\sqrt{2U}\right)
^{2}+\left\vert \mathcal{D}_{\pm}\phi\right\vert ^{2}\pm\partial_{i}%
\mathbb{J}_{i}\right. \nonumber\\[-0.15cm]
& \\[-0.15cm]
&  ~\ \ \ \ \left.  \frac{{}}{{}}\pm B\left(  \sqrt{2U}+e\left\vert
\phi\right\vert ^{2}+\xi_{j}\partial_{j}\left\vert \phi\right\vert
^{2}\right)  \right]  .\nonumber
\end{align}

At this point, we choose the interaction $U$ with the purpose of achieving self-dual or first-order differential equations. The interaction satisfying this requirement is
\begin{equation}
{U\left(  \left\vert \phi\right\vert ,\partial_{\mu}\left\vert \phi\right\vert
\right)  }=\frac{1}{2}\left(  ev^{2}-e\left\vert \phi\right\vert ^{2}-\xi
_{j}\partial_{j}\left\vert \phi\right\vert ^{2}\right)  ^{2}.\label{pot_XX}%
\end{equation}
We see that the Lorentz violation induces the presence of derivative terms in the interaction providing self-dual configurations. This pattern was already observed in other Maxwell--Higgs models supporting Lorentz violation
\cite{Lazar,Miller2}.

Then, the energy may be rewritten as follows
\begin{align}
E &  =\int d^{2}x\left[  \frac{1}{2}\left[  B\mp\left(  ev^{2}-e\left\vert
\phi\right\vert ^{2}-\xi_{j}\partial_{j}\left\vert \phi\right\vert
^{2}\right)  \right]  ^{2}\right. \nonumber\\[-0.15cm]
& \\[-0.15cm]
&  ~\ \ \ \ \left.  \frac{{}}{{}}+\left\vert \mathcal{D}_{\pm}\phi\right\vert
^{2}\pm ev^{2}B\pm\partial_{i}\mathbb{J}_{i}\right]  .\nonumber
\end{align}
Under appropriate boundary conditions, the contribution stemming from the term $\partial_{i}\mathbb{J}_{i}$ to the total energy is null. In such a way, the energy possesses a lower bound or BPS limit%
\begin{equation}
E\geq\pm ev^{2}\int d^{2}xB=\pm ev^{2}\Phi,\label{energy_BPS}%
\end{equation}
which relates the total energy proportional to the magnetic flux.
This bound is saturated by the fields fulfilling the BPS or self-dual equations,
\begin{align}
&  \displaystyle{\mathcal{D}_{\pm}\phi=0,}\label{bsp_1}\\[0.2cm]
&  \displaystyle{B=\pm\left(  ev^{2}-e\left\vert \phi\right\vert ^{2}-\xi
_{j}\partial_{j}\left\vert \phi\right\vert ^{2}\right)  .}\label{bsp_2}%
\end{align}
We can see that the second BPS equation depends of the LV vector background (magnitude and direction). Similar dependence has been also reported for other LV electrically charged configurations \cite{Miller2,Lazar,claudio1}. However, there is a difference between these cases: in the present model, the Higgs field and magnetic field solutions for
$-\xi_{j}$ are different from the ones engendered by $\xi_{j}$. On the other hand, in the the electrically charged solutions of Ref. \cite{Miller2,Lazar,claudio1}, when the fixed LV background, $V_{i}$, is changed to $-V_{i}$, no modifications are reported for the Higgs field and magnetic field solutions, while the electric sector is affected: the scalar potential $A_{0}$ changes to $-A_{0}$ for $-V_{i}$, that is, the direction of the electric field is inverted.

Finally, we affirm that the first-order differential equations
reproduce the second order Euler-Lagrange equations given by Eqs. (\ref{ampere}) and (\ref{higgs}), with the interaction $U$ given by Eq.
(\ref{pot_XX}).

\section{The self-dual vortex solutions}

In {the} present section, the purpose is to study self-dual vortex solutions of the
BPS equations (\ref{bsp_1}) and (\ref{bsp_2}). The vortices are naturally described in polar coordinates $(r,\theta)$ by using the following
\textit{Ansatz:}
\begin{align}
{\phi}&=vg(r)e^{in\theta}, & A_{\theta}&=-\frac{a(r)-n}{er}, & A_{r}&=0,\label{Ans}%
\end{align}
where $n=\pm1,\pm2,\pm3{,\ldots}$ is the winding number characterizing the vortex
solution. The \textit{Ansatz} provides a simple expression for the magnetic field,
\begin{equation}
B(r)=-\frac{a^{\prime}}{er},\label{magnetic_x}%
\end{equation}
with ${}^{\prime}\equiv d/dr$, a derivative in relation to the variable $r$. The profile functions $g(r)$ and $a(r)$ are well behaved functions satisfying the boundary conditions%
\begin{align}
g(0) &  =0, & a(0)=n,\label{bc00}\\[0mm]
g(\infty) &  =1, & a(\infty)=0,\label{bc01}%
\end{align}
providing finite energy configurations. In the next
subsection (\ref{BBCC}), we will explicitly show the compatibility of them with the BPS equations.

Now, we come back to the BPS energy (\ref{energy_BPS}), which can be easily computed by using the expression (\ref{magnetic_x}) for the magnetic field, that is%
\begin{equation}
E_{bps}=2\pi v^{2}\left\vert n\right\vert,\label{XXX}%
\end{equation}
which reveals that the energy is quantized, i.~e., proportional to $n$, the winding number characterizing the vortex solution.

In the \textit{Ansatz} (\ref{Ans}), {both} the current conservation $\partial_{k}{J}_{k}=0$ and the BPS equation (\ref{bsp_1}) {provide}
\begin{equation}
\xi_{\theta}=0,
\end{equation}
as a consistency condition. Thus, the BPS equations (\ref{bsp_1})--(\ref{bsp_2}) for the self-dual vortices become
\begin{align}
&  \displaystyle{g^{\prime}=\pm g\left(  \frac{a}{r}-\xi_{r}\frac{a^{\prime}%
}{er}\right)  ,}\label{bsp_1.1}\\[0.2cm]
&  \displaystyle{B=-\frac{a^{\prime}}{er}=\pm ev^{2}\left(  1-g^{2}\right)
\mp2v^{2}\xi_{r}gg^{\prime},}\label{bsp_2.1}%
\end{align}
with the upper (lower) signal {standing} for $n>0$ ($n<0$). Such as pointed out in the previous section, the solutions $g(r)$ and $a(r)$ will be different for $\xi_{r}$ and $-\xi_{r}$. However, the connection between the solutions for $n>0$ and $n<0$ is maintained: $g^{<}(r)=g^{>}(r)$ and $a^{<}(r)=-a^{>}(r)$, independently of the $\xi_{r}$ signal.

The BPS energy density for the vortices, in the Ansatz (\ref{Ans}), obtained from Eq. (\ref{eenn}), is
\begin{equation}
{\varepsilon}_{bps}=B^{2}+2v^{2}g^{2}\left( \frac{a}{r}+\xi_{r}B\right)^{2}.
\end{equation}
It is positive-definite for all values of the Lorentz-violating parameter, $\xi_{r}$.

\subsection{Checking the boundary conditions}

\label{BBCC}

We proceed to check if the boundary conditions (\ref{bc00}) and (\ref{bc01}) are compatible with the BPS equations (\ref{bsp_1.1}) and (\ref{bsp_2.1}) in its dimensionless form. The behavior of the solutions near to the origin ($r\rightarrow0$) is computed by using power-series method, yielding
\begin{align}
g(r) & =G_{n}r^{n}+\xi_{r}ev^{2}G_{n}r^{n+1}+{\cdots},\label{bc0_g}\\[0.2cm]
a(r) & =n-\frac{e^{2}v^{2}}{2}r^{2}+\frac{2nev^{2}}{2n+1}\xi_{r}(G_{n})^{2}r^{2n+1}+{\cdots},
\end{align}
where $G_{n}$ is a fixed constant for every $n$, determined numerically. These expansions confirm the boundary conditions imposed in (\ref{bc00}).

For $r\rightarrow\infty$, we are looking for field profiles whose behavior is similar to Abrikosov--Nielsen--Olesen's ones, so we attain
\begin{align}
g(r) &  \sim1-G_{_{\!\infty}}r^{-1/2}e^{-\beta r}, \nonumber\\[-0.1cm]
     & \label{Inf_1}\\[-0.1cm]
a(r) &  \sim\frac{2G_{_{\!\infty}}\beta e}{e+\beta{\xi_{r}}}r^{1/2}e^{-\beta r},\nonumber
\end{align}
where $G_{_{\!\infty}}$ can be determined numerically and $\beta$, the mass of the bosonic fields, is given by
\begin{equation}
\beta=\frac{\sqrt{2}ev}{\sqrt{1+2v^{2}\left({\xi_{r}}\right)^{2}}}.\label{beta}%
\end{equation}
We observe that Maxwell--Higgs's scale is recuperated when Lorentz-violating
parameter, $\xi_{r},$ is null, i.~e., ${\xi_{r}}=0$.

\subsection{Numerical analysis}

For such a purpose, we do the rescaling $\rho\rightarrow evr$ and the following redefinitions:
\begin{align}
g(r)&\rightarrow g(\rho),         &  a(r)&\rightarrow a(\rho), \nonumber \\[3mm]
B(r)&\rightarrow ev^{2}B(\rho),   &  \varepsilon_{bps}(r)&\rightarrow v^{2}{\varepsilon}_{bps}(\rho),\\[2mm]
\xi_{r}&\rightarrow\frac{\delta}{v},  \nonumber
\end{align}
{which} lead to the dimensionless version of the BPS equations,
\begin{align}
g^{\prime}&=\pm g\left(\frac{a}{\rho}-\delta \frac{a^{\prime}}{\rho}\right),\label{bq1}\\[1mm]
B&=-\frac{a^{\prime}}{\rho}=\pm\left(  1-g^{2}\right)\mp2\delta gg^{\prime}.\label{bq2}%
\end{align}
Hereafter, the parameter $\delta$ stands for the Lorentz-violating contributions.

In the sequel, we perform a numerical analysis of the solutions of the BPS
equations (\ref{bq1}) and (\ref{bq2}) in two situations: in the first one
considers $n=1$ fixed and some values of $\delta$ in the interval $[-1,1]$; in
the second case, we have compare the profiles for $\delta=-0.75,0,0.75$ and
several values of the winding number $n$. In every case, we describe and
highlight the modifications introduced by the Lorentz-violating parameter
($\delta\neq0)$ when compared to the usual MH model ($\delta=0$).

\subsubsection{Numerical solutions $n=1$ and $-1\leq\delta\leq1$}

Fig. \ref{higgs_p} shows the behavior of the Higgs field profiles for
$n=1$ and some values of $\delta$. When compared to the MH ones ($\delta=0$),
they are narrower for more negative values of $\delta$ and wider for more
positive values of $\delta$. In both cases, far away from the origin, the
larger is $|\delta|$, more slowly the profiles converge to the vacuum value in
relation to the MH one, in accordance with the mass scale defined in Eq.
(\ref{beta}).

\begin{figure}
\centering
\scalebox{0.9}{\includegraphics[width=8.5cm]{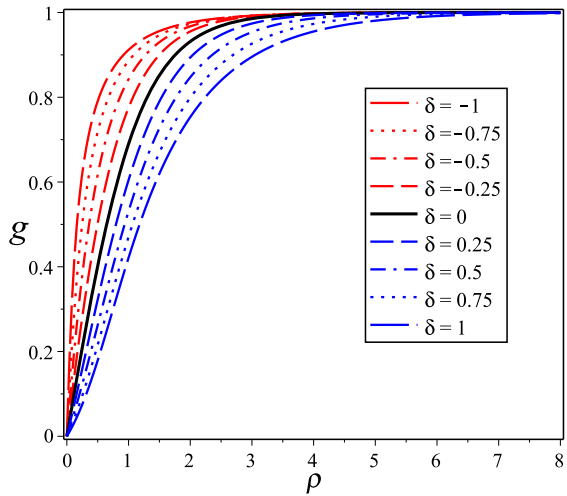}}\caption{The Higgs
field profile ${g}(\rho)$ for $n=1$. The red lines represent the solutions for
$\delta<0$, the black line ($\delta=0$) gives the BPS solution for the MH
model, and the blue lines for $\delta>0$.}%
\label{higgs_p}%
\end{figure}

\begin{figure}
\centering
\scalebox{0.9}{\includegraphics[width=4.45cm]{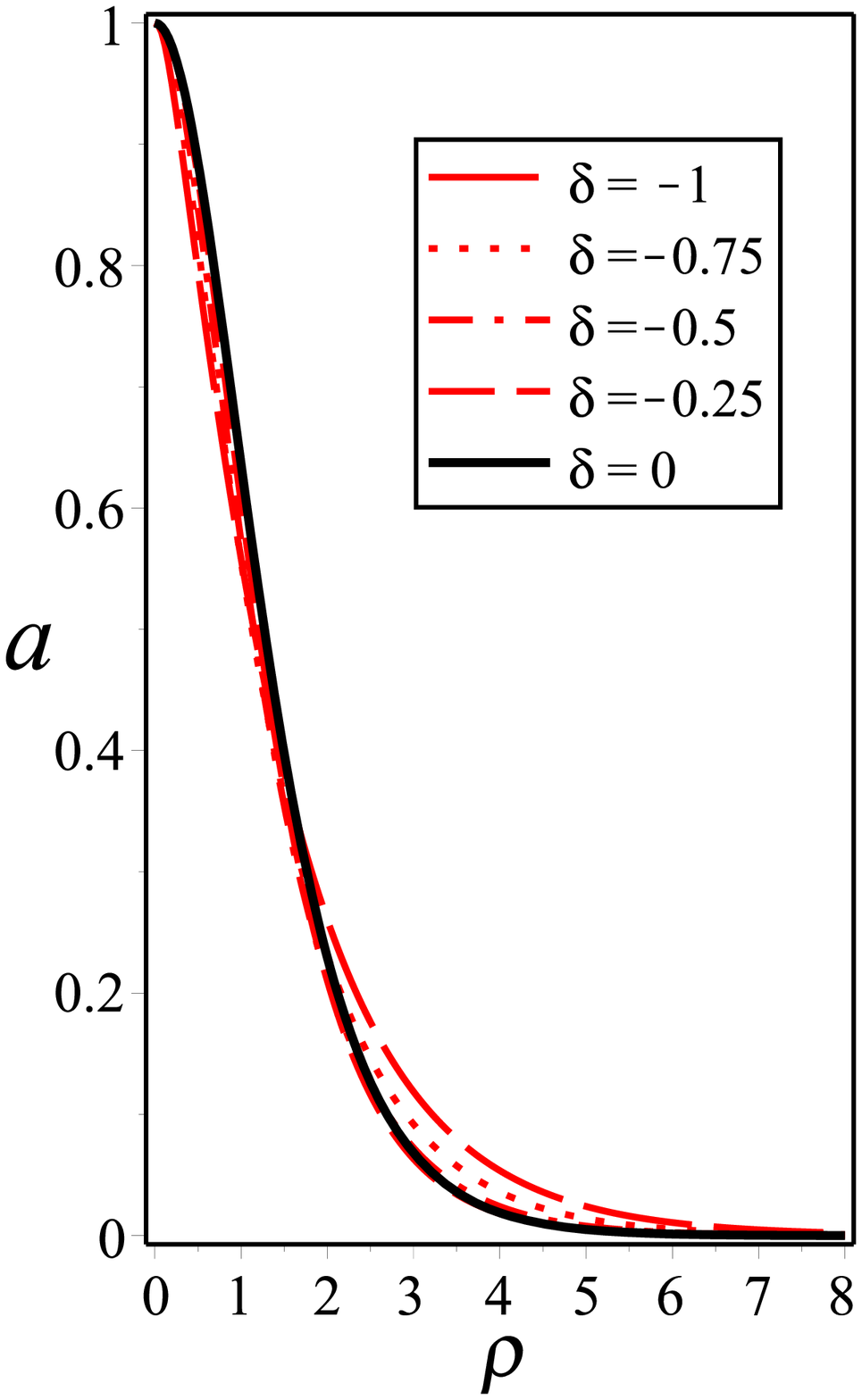}\!
\includegraphics[width=4.18cm]{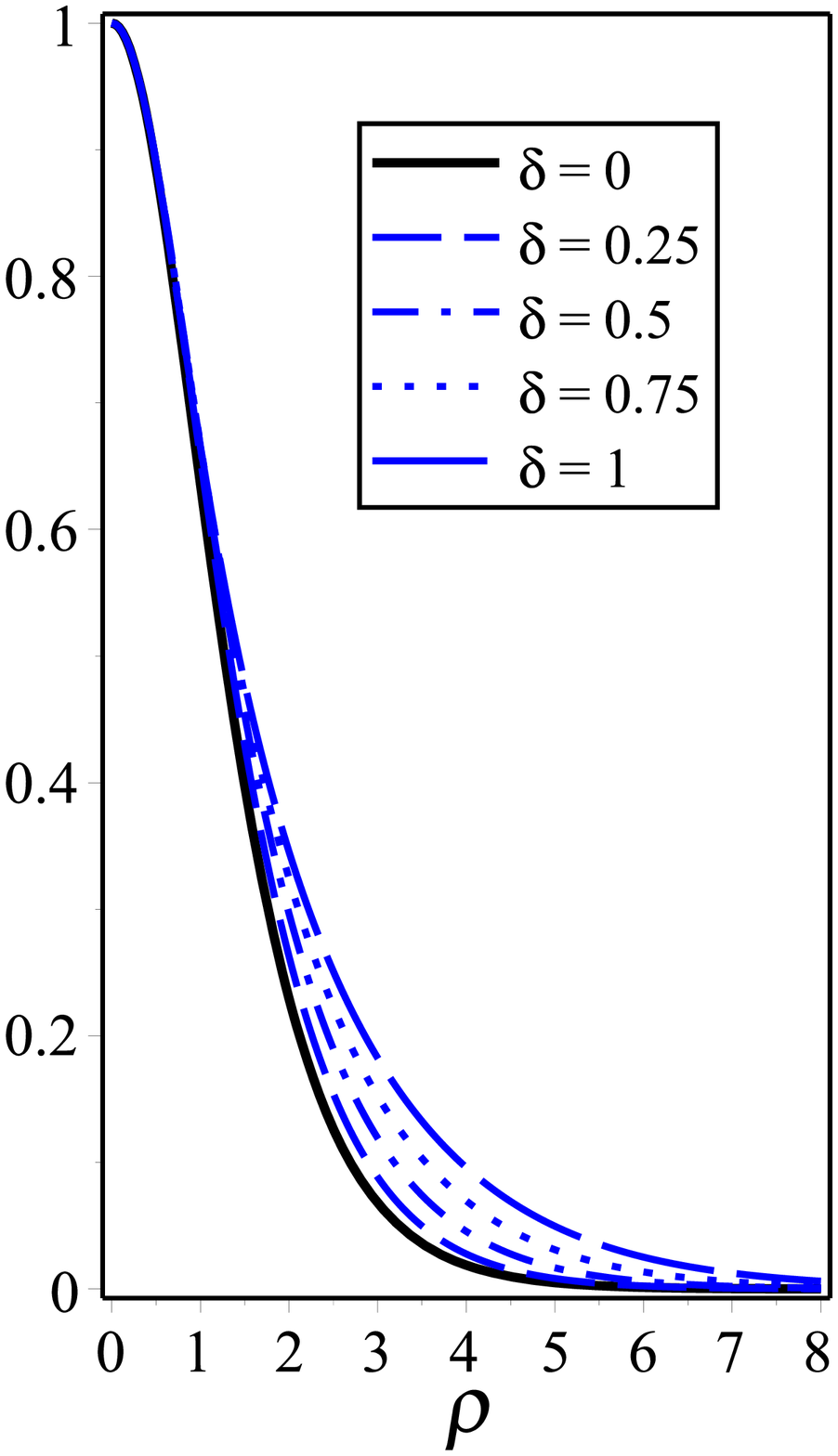}}\caption{The gauge field profile
${a}(\rho)$ for $n=1$. The red lines represent the solutions for $\delta<0$,
the black line ($\delta=0$) gives the BPS solution for the MH model, and the
blue lines for $\delta>0$.}%
\label{gauge_p}%
\end{figure}

The left-side of Fig. \ref{gauge_p} shows the gauge field profiles $a(\rho)$
for $n=1$ and $\delta<0$, which near the origin become slightly narrower than
the MH ones ($\delta=0$). The right-side of Fig. \ref{gauge_p} shows the
profiles for $\delta>0$. In this case, near the origin they are almost
overlapped to the ones of the MH model. We observe that for $\delta\neq0$, at
some distance from the origin, the profiles become wider than the MH one. This
effect is minor for $\delta<0$\ and major for $\delta>0$. This deviation
augments with increasing $\left\vert \delta\right\vert $ values.

\begin{figure}
\centering
\scalebox{0.9}{\includegraphics[width=8.5cm]{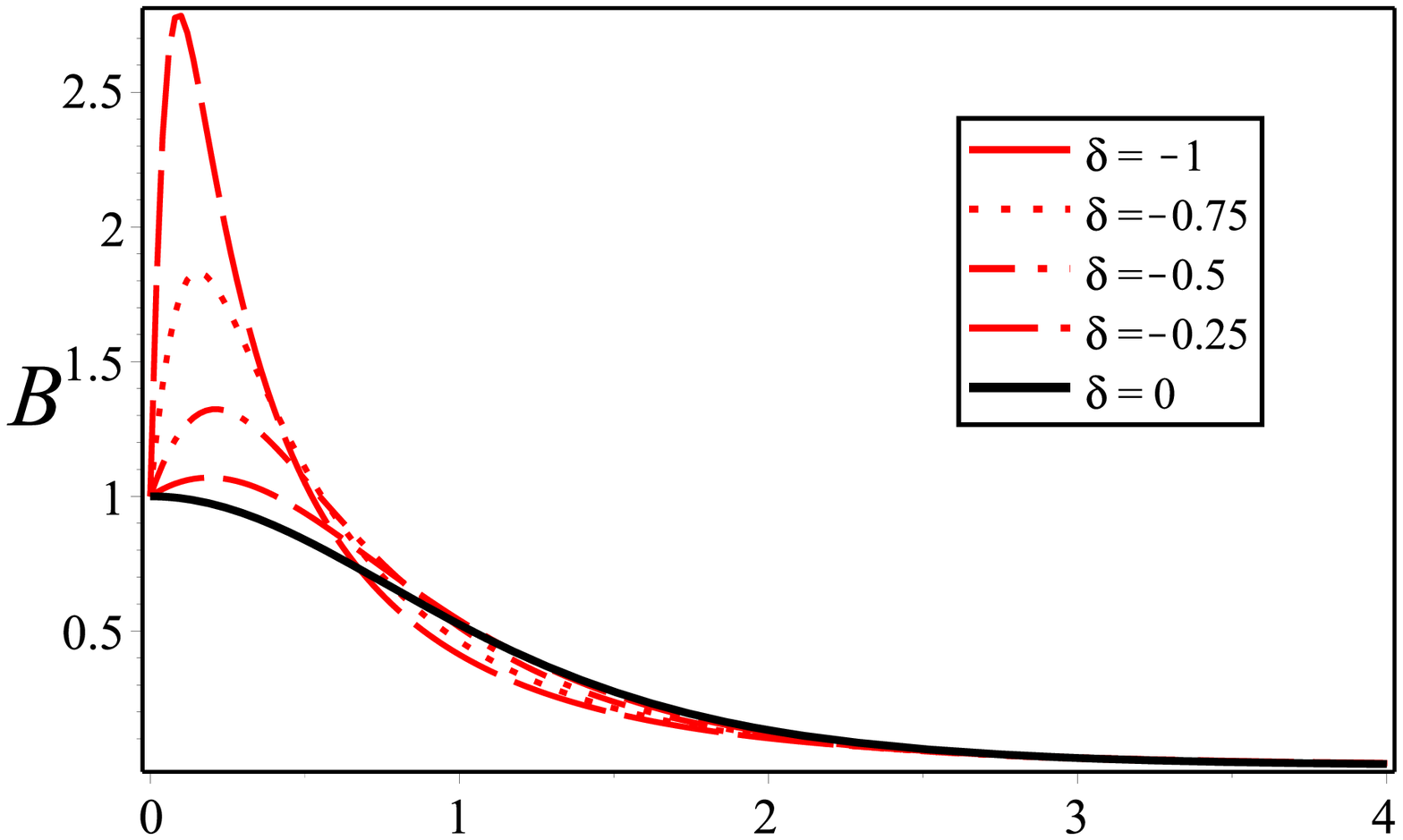}}
\scalebox{0.9}{\includegraphics[width=8.5cm]{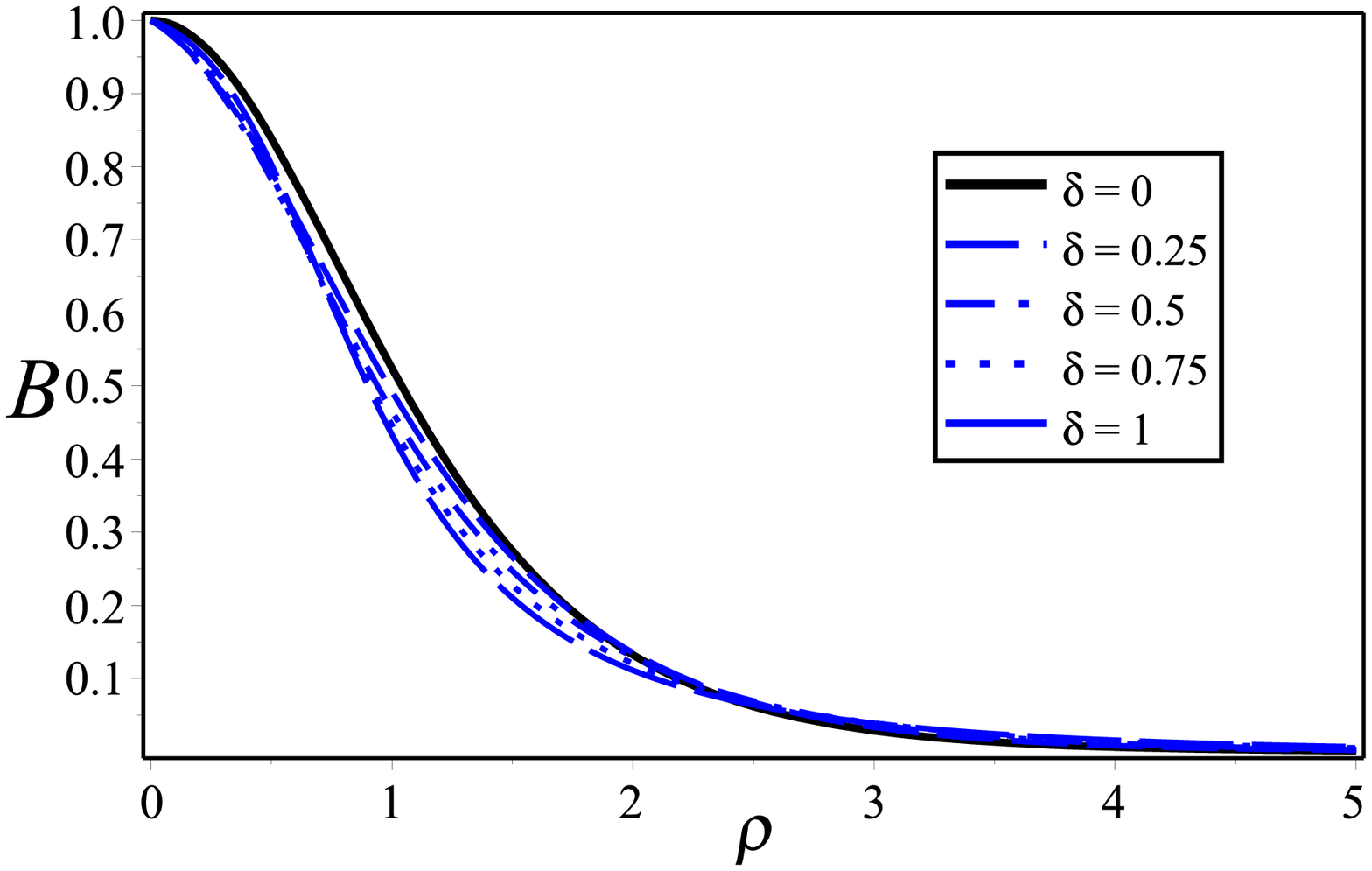}}\caption{ The
magnetic field profile $B(\rho)$ for $n=1$. The red lines represent the
solutions for $\delta<0$, the black line ($\delta=0$) gives the BPS solution
for the MH model, and the blue line for $\delta>0$.}%
\label{magnetic_p}%
\end{figure}

Upper Fig. \ref{magnetic_p} shows the magnetic field profiles, $B(\rho)$, for
$n=1$ and $\delta<0$. At the origin, the magnetic field presents a finite
value, $ev^{2},$ as the MH one. Close to the origin the LV parameter plays
great influence, yielding a peak amplitude that forms a ring-like shaped
structure. For more negative values of $\delta$, the peak is higher and more
localized, while it becomes lower and wider as $\left\vert \delta\right\vert $
diminishes. Far from the origin, the magnetic field decays as much as the MH
solution. This ring-like behavior, obtained for $n=1 $, differs greatly from
the lump-like MH ones, resembling the profiles of the Chern--Simons--Higgs
models. Lower Fig. \ref{magnetic_p} shows the magnetic field profiles for
$\delta>0$, which are very similar lumps to the MH ones, with the same value
$ev^{2}$ at the origin. As one moves from the origin, the amplitude of
$B(\rho)$ becomes slightly lesser than the MH one, yielding a little more
localized {defect.}

\begin{figure}
\centering
\scalebox{0.9}{\includegraphics[width=8.5cm]{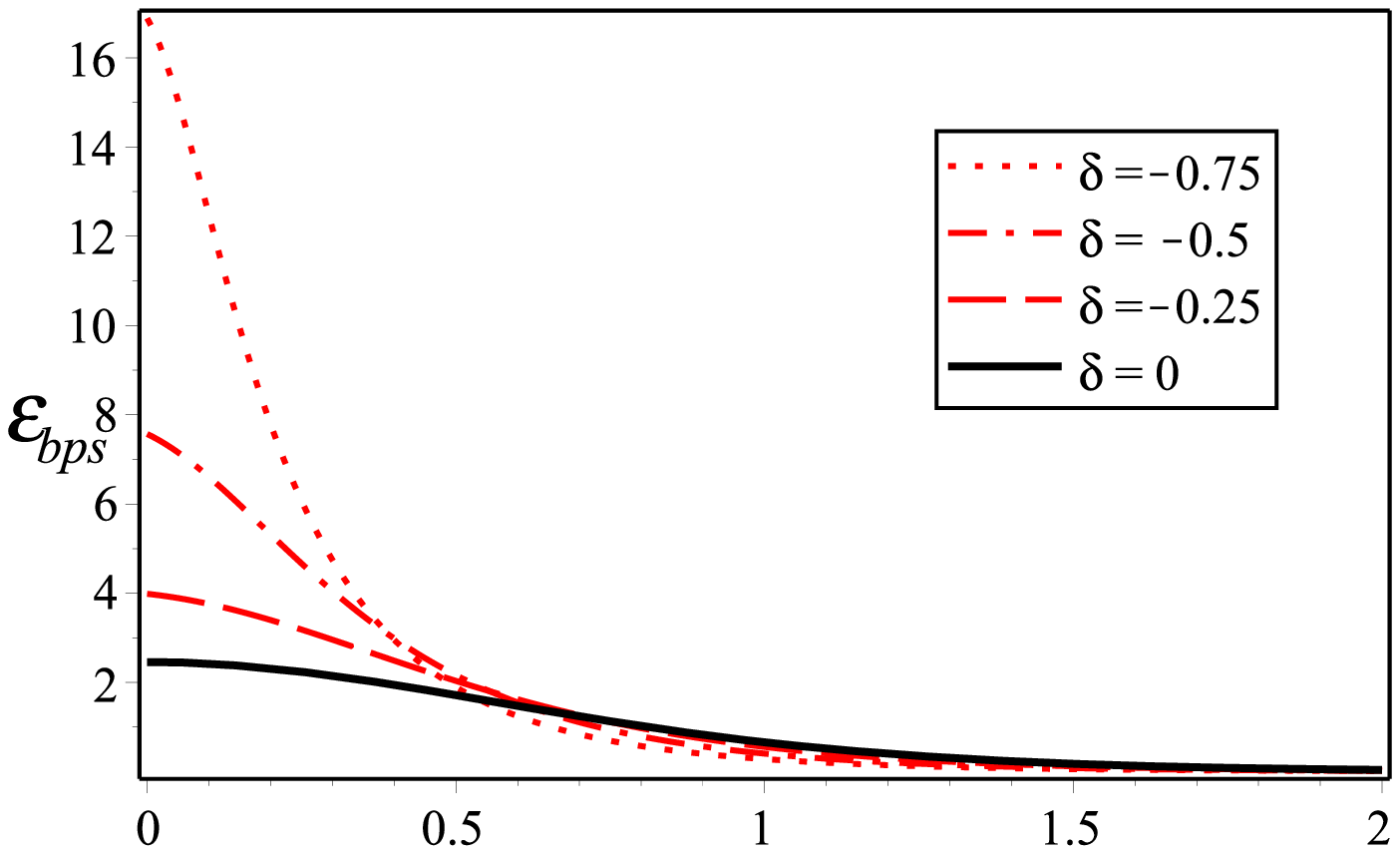}}
\scalebox{0.9}{\includegraphics[width=8.5cm]{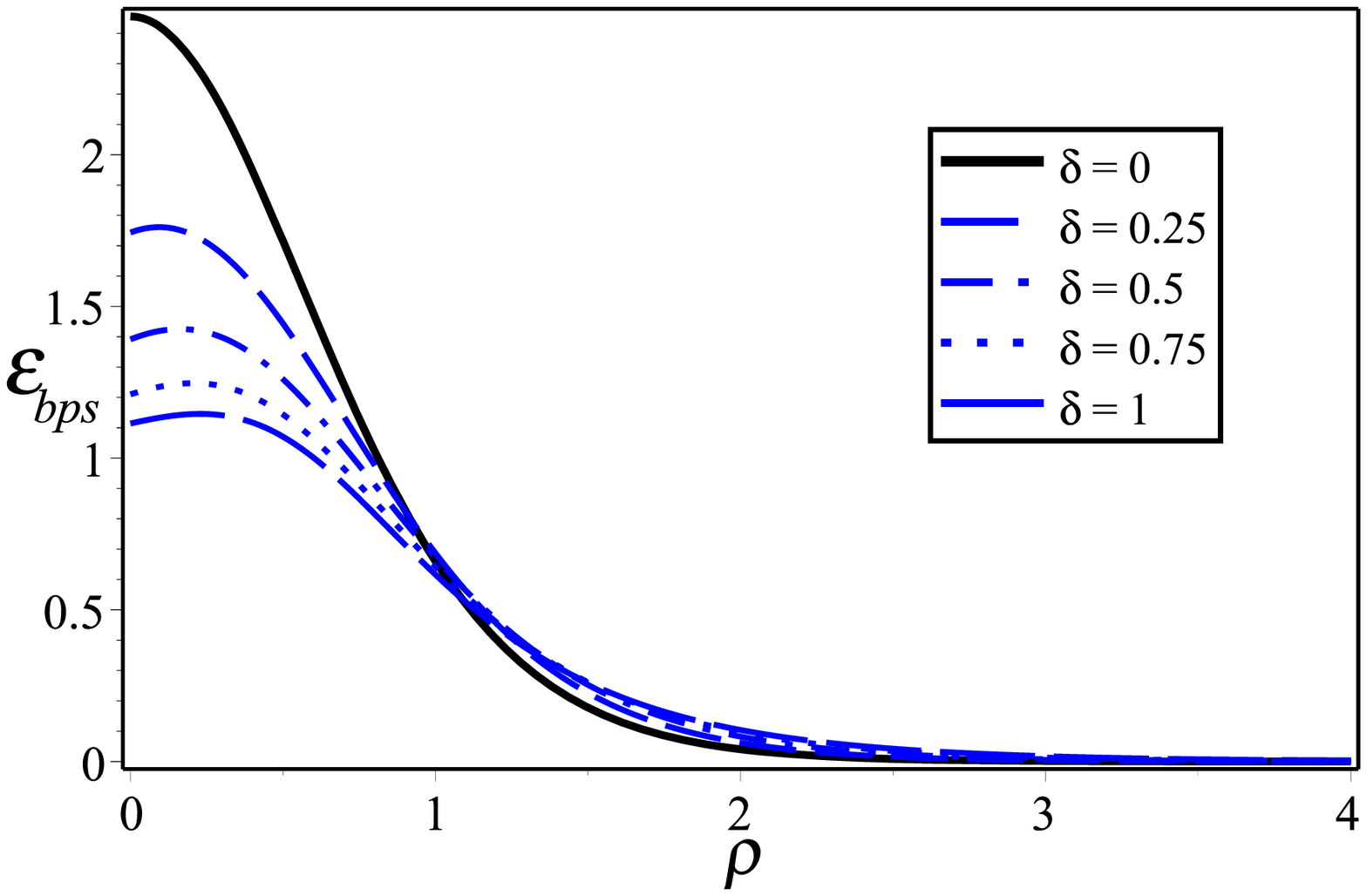}}\caption{The BPS
energy density profile $\varepsilon_{bps}(\rho)$ for $n=1$. The red lines
represent the solutions for $\delta<0 $, the black line ($\delta=0$) gives the
BPS solution for the MH model, and the blue lines for $\delta>0$.}%
\label{energia_p}%
\end{figure}

Fig. \ref{energia_p} shows the profiles of the BPS energy density
$\varepsilon_{bps}$ for $n=1$. For $\delta<0$, we observe that the
$\varepsilon_{bps}$ profiles present a lump shape similarly to the MH one,
with a more pronounced and localized peak near to the origin. The peak
amplitude and localization {increase} when $\delta$ becomes more negative. On
the other hand, for $\delta>0$ (lower figure), the profiles maintain the
lump-like shape, with the peak amplitude becoming smaller and less localized
than the MH one. The peaks move away from the origin as $\delta$ increases.
Numerically it is observed that for very large positive values of $\delta$,
the BPS energy density amplitude at the origin has its lower bound at the
value $e^{2}v^{4}$.

\subsubsection{Numerical solutions for $\delta=-0.75,0,0.75$ and
$n=1,2,4,6,10,20$}

{Figs.} \ref{higssX} and \ref{gaugex} show the behavior of Higgs and gauge
field profiles for various $n$ values with $\delta$ negative, null and
positive. In both cases, the profiles become wider with raising $n$\ values,
but maintain a behavior similar to the case $n=1$ exhibited in Figs.
\ref{higgs_p} and \ref{gauge_p}, respectively.

\begin{figure}[ptb]
\centering
\scalebox{0.9}{\includegraphics[width=8.5cm]{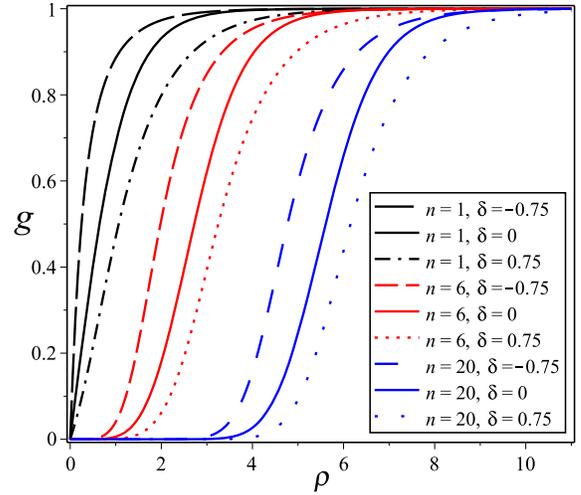}}\caption{The Higgs
field profiles $g(\rho)$ for $n=1,6,20$ and $\delta=-0.75,0,0.75$. The solid
lines, $\delta=0$, represent the BPS solutions for the MH model.}%
\label{higssX}%
\end{figure}

\begin{figure}
\centering
\scalebox{0.9}{\includegraphics[width=8.3cm]{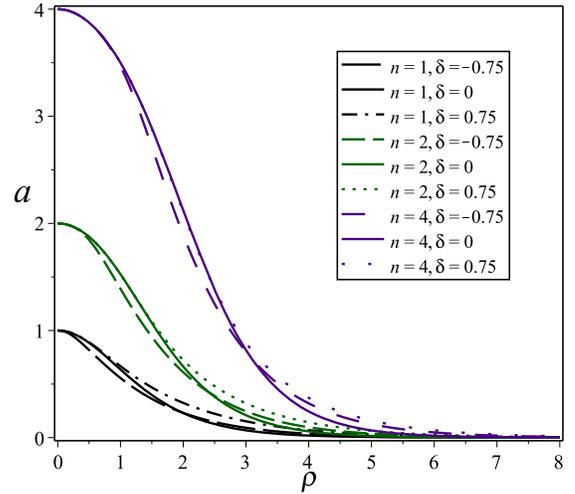}}\caption{The gauge
field profiles $a(\rho)$ for $n=1,2,4$ and $\delta=-0.75,0,0.75$. The solid
lines, $\delta=0$, represent the BPS solutions for the MH model.}%
\label{gaugex}%
\end{figure}

Fig. \ref{magneticox} depicts the magnetic field profiles for some
values of $n$ and $\delta$. The upper figure provides the profiles for a
negative value of $\delta$ ($\delta=-0.75$). It is observed the presence of
peaks, corresponding to a ring-like behavior, which are more accentuated and
closer to the origin for small winding number values. For large values of $n$,
the ring structures have maximum amplitudes progressively smaller and
located at an increasing distance from the origin. This behavior {contrasts} with the Maxwell--Higgs one, where the magnetic field profile displays
a plateau whose width increases for larger $n$ values. So, the nonminimal
coupling induces a ring-like behavior for the Abelian Higgs vortices when the
LV parameter ($\delta$) takes negative values. The lower figure represents the
magnetic field for a positive value of $\delta$ ($\delta=0.75$). It is
observed that, for all values of $n$, the profiles follow closely the
Maxwell--Higgs magnetic field behavior, presenting only a tiny deviation.

\begin{figure}
\centering
\scalebox{0.9}{\includegraphics[width=8.5cm]{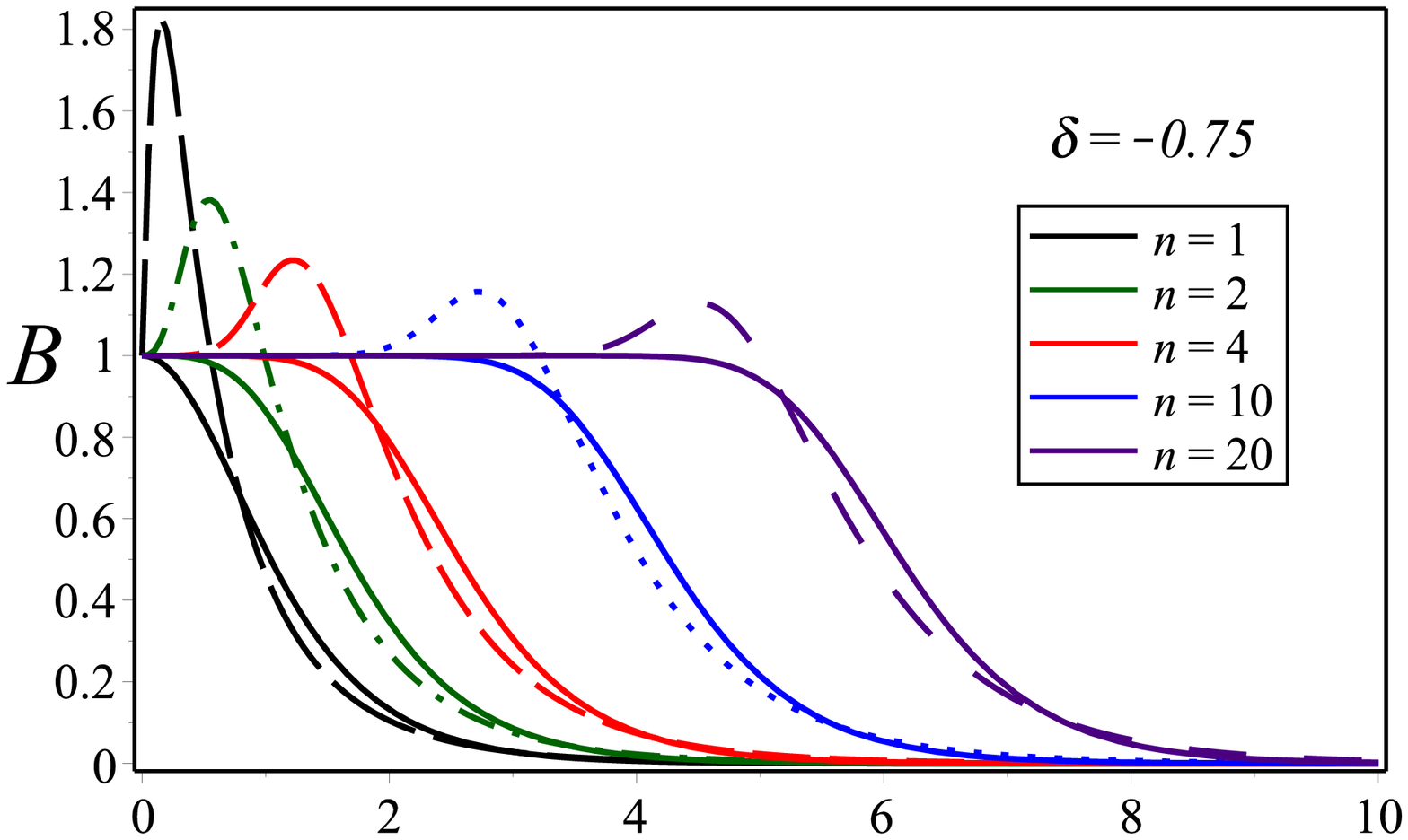}}
\scalebox{0.9}{\includegraphics[width=8.5cm]{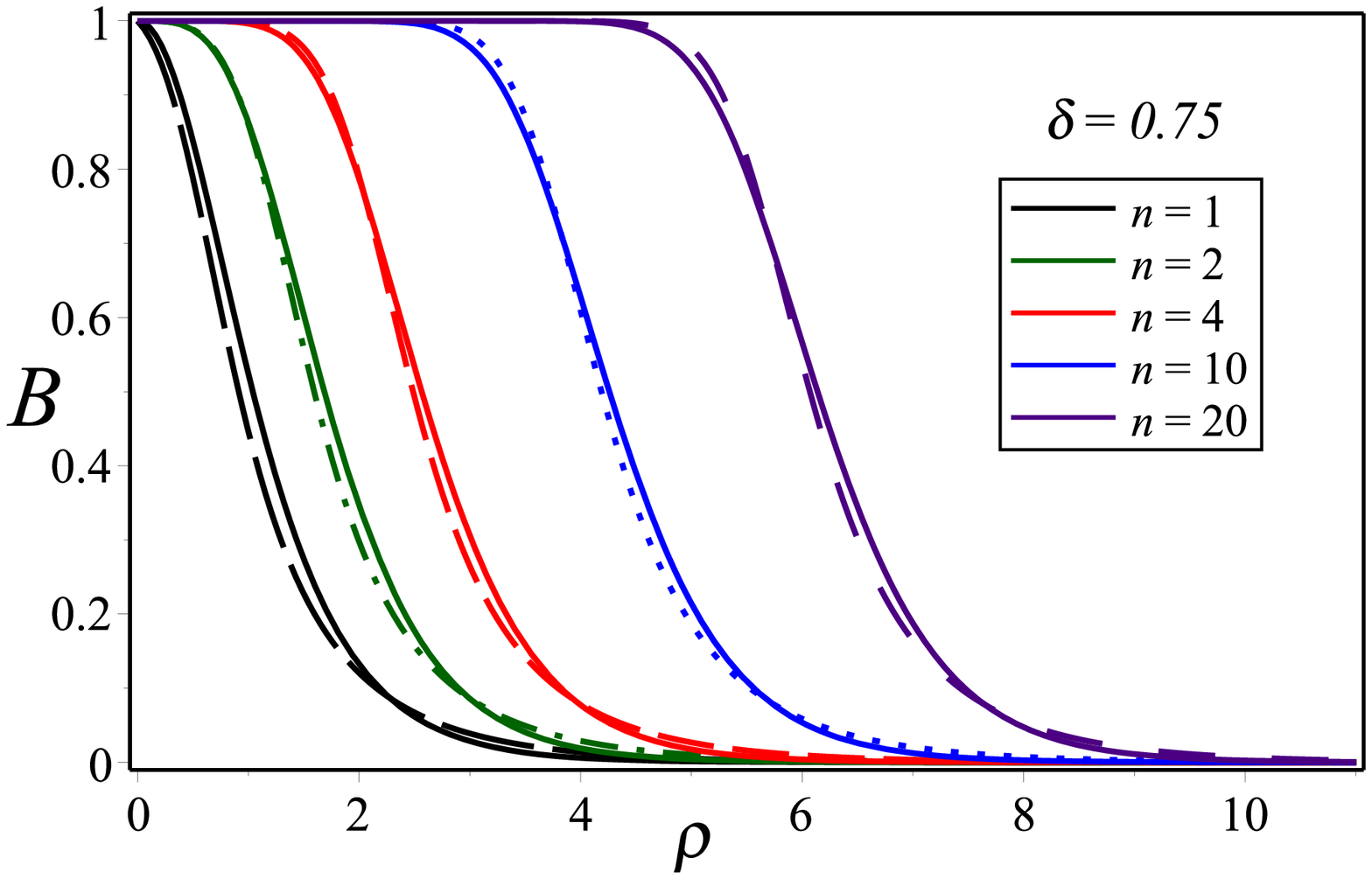}}\caption{The magnetic
field profiles $B(\rho)$ for $\delta=-0.75$ (upper figure) and $\delta=0.75$
(lower figure). In both cases, $n=1,2,4,10,20$. The solid lines represent the
BPS solutions for the MH model.}%
\label{magneticox}%
\end{figure}

\begin{figure}
\centering
\scalebox{0.9}{\includegraphics[width=8.5cm]{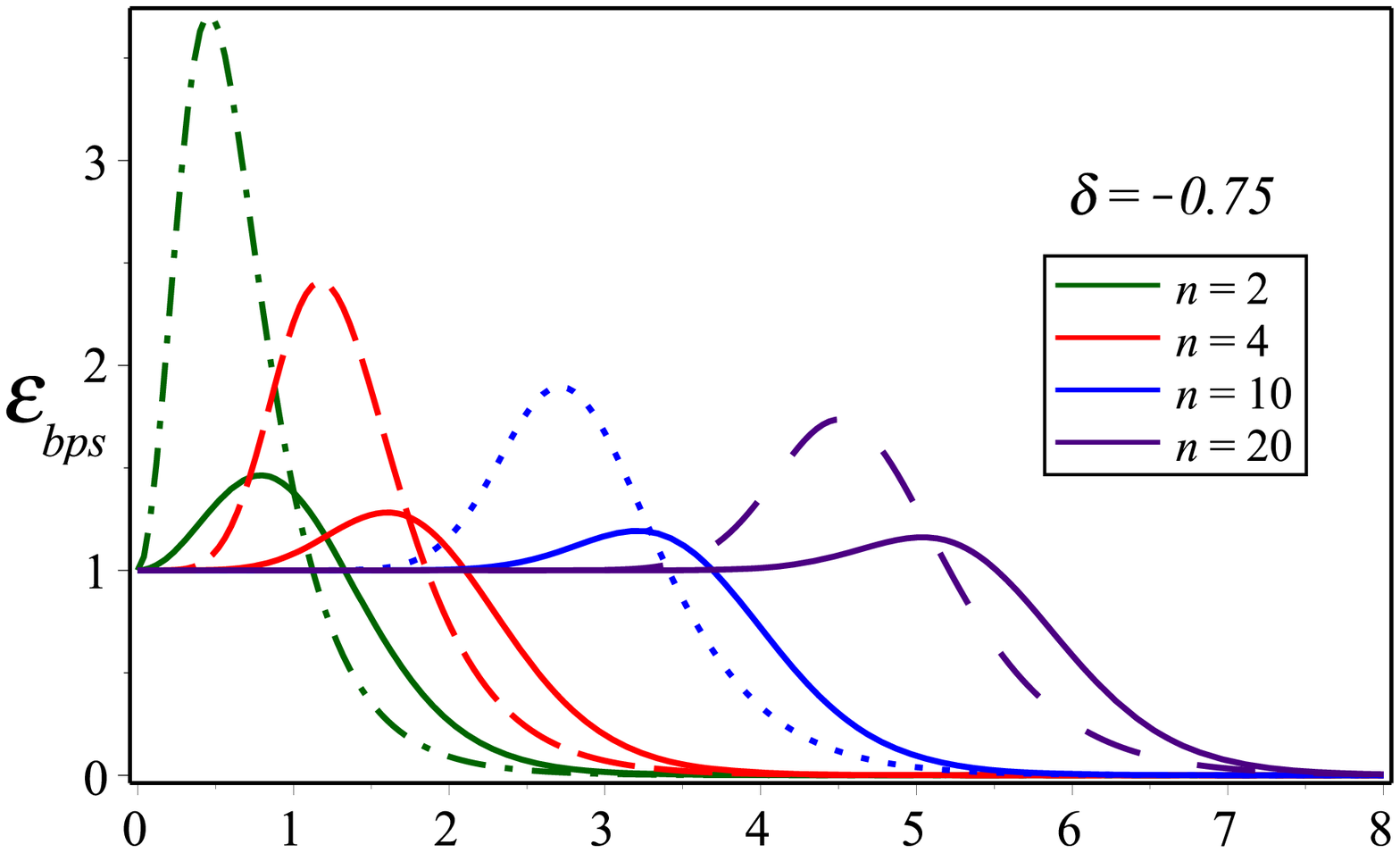}}
\scalebox{0.9}{\includegraphics[width=8.5cm]{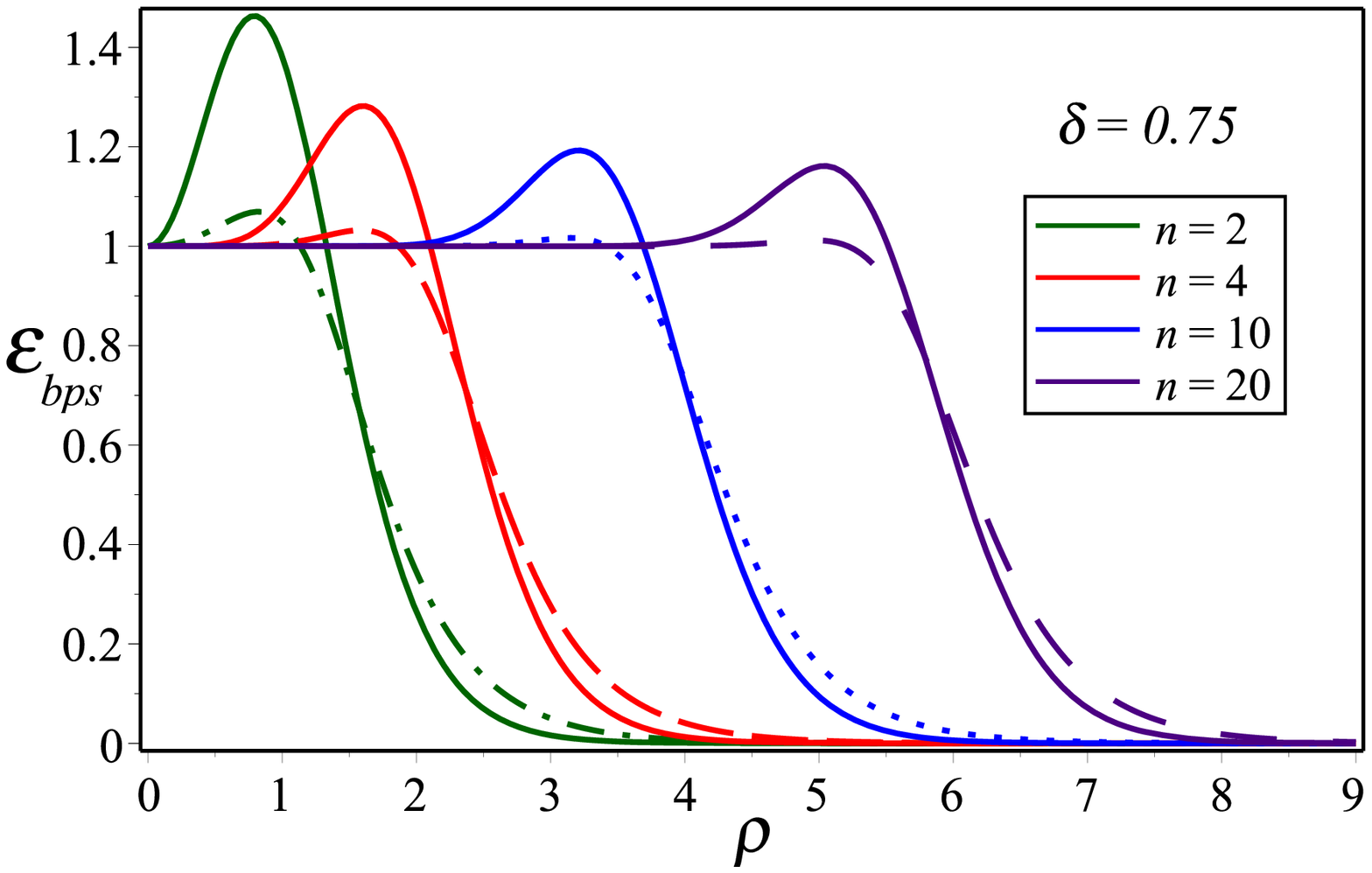}}\caption{BPS energy
density profiles $\varepsilon_{bps}(\rho) $ for $\delta=-0.75$ (upper figure)
and $\delta=0.75$ (lower figure). In both cases, $n=2,4,10,20$. The solid
lines represent the BPS solutions for the MH model.}%
\label{energiax}%
\end{figure}

Fig. \ref{energiax} depicts the BPS energy density profiles for some
values of $n\geq2$ and fixed $\delta$. Despite the presence of the Lorentz
violation, the profiles also display the ring-like behavior, as those of the
MH model. In the upper picture, a negative LV parameter value ($\delta=-0.75$)
enhances the amplitude peak of the ring shaped profiles, which become
progressively lower for larger values of the winding number $n$. On the other
side, in the lower figure, one notices that a positive LV parameter
($\delta=0.75)$ plays an opposite effect on the solutions, turning the peak
amplitude much lesser than the MH ones. Consequently, for sufficiently large
values of $n$, the Lorentz violation makes disappear the ring-like structure,
reducing it to a simple plateau whose width increases with raising $n$ values.

\section{Remarks and conclusions}

We have shown the existence of topological BPS or self-dual solutions in a MH
model endowed with a CPT-odd and Lorentz-violating nonminimal coupling between
the gauge and the Higgs fields. The Lorentz violation modifies self-dual or
BPS equations of the MH model. Specifically, the Lorentz violation changes the
usual symmetry breaking $|\phi|^{4}$-potential by introducing a new derivative
self-interaction whose coupling constant is the own LV vector background. The present CPT-odd self-dual configurations are now electrically
neutral in contraposition to the CPT-odd charged cases in which the Lorentz
violation is included in the kinetic sector \cite{Lazar,Alyson,claudio2}. Another
feature of this nonminimal model is that the Higgs field and magnetic field
solutions for $-\xi_{j}$ are different from the ones for $\xi_{j}$.

We have solved the BPS equation for axially symmetric vortex solutions. Besides the control on the width of the vortex core, the magnetic field profiles undergo relevant modifications for negative values of the LV parameter, inducing a ring-like behavior similar to one appearing in Abelian
Higgs models containing the Chern--Simons term. The BPS energy density also
suffers significant deviations from the MH profiles for all values of the LV
parameter, being observed that, for $n\geq2$, the ring-like behavior can be
enhanced (becoming much closer to the MCSH behavior) for a positive LV
parameter or strongly attenuated (in relation to the MH typical profiles) for
negative values of LV parameter. We have thus argued that the
consideration of a preferred direction in spacetime, by means of a nonminimal
coupling, is a factor that can indeed enrich the description of vortex structures.

\begin{acknowledgments}
We thank CAPES, CNPq/483863/2013-0 and FAPEMA/UNIVERSAL-00782/15 (Brazilian agencies) for partial financial support.
\end{acknowledgments}

\hfil


\begin{thebibliography}{99}                                                                                               %
\bibitem {Abrikosov}A. A. Abrikosov, Zh. Eksp. Teor. Fiz. 32, 1442 (1957);
Sov. Phys. - JETP 5, 1174 (1957); J. Phys. Chem. Solid. 2, 199 (1957).

\bibitem {GLT}V. L. Ginzburg and L. D. Landau, JETP 20, 1064 (1950).

\bibitem {Onsager}L. Onsager, Nuovo Cimento 9, 6, 279 (1949); R. P. Feyn- man,
in Chapter II Application of Quantum Mechanics to Liquid Helium, Progress in
Low Temperature Physics, Vol. 1, edited by C. Gorte (Elsevier, 1955) pp. 1753.

\bibitem {NOlesen}H. Nielsen and P. Olesen, Nucl. Phys. B 61, 45 (1973).

\bibitem {Vega}F. A. Schaposnik and H. J. de Vega, Phys. Rev. D 14, 1100 (1976).

\bibitem {Hyun}S. Hyun, J. Shin, J. H. Yee, and H.-j. Lee, Phys. Rev. D 55,
3900 (1997).

\bibitem {SV}H. J. de Vega and F. A. Schaposnik, Phys. Rev. Lett. 56, 2564
(1986); S. K. Paul and A. Khare, Phys. Lett. B 174, 420 (1986).

\bibitem {HKJ}J. Hong, Y. Kim, and P. Y. Pac, Phys. Rev. Lett. 64, 2230
(1990); R. Jackiw and E. J. Weinberg, 64, 2234 (1990); R. Jackiw, K. Lee, and
E. J. Weinberg, Phys. Rev. D 42, 3488 (1990); A. Khare, Phys. Lett. B 255, 393
(1991); Proc. Indian Natn. Sci. Acad. A 61, 161 (1995).

\bibitem {BPS}M. K. Prasad and C. M. Sommerfeld, Phys. Rev. Lett. 35, 760
(1975); E. B. Bogomol'nyi, Yad. Fiz. 24, 861 (1976); Sov. J. Nucl. Phys. 24,
449 (1976).

\bibitem {SME1}D. Colladay and V. A. Kosteleck\'{y}, Phys. Rev. D 55, 6760
(1997); 58, 116002 (1998); S. Coleman and S. L. Glashow, 59, 116008 (1999).

\bibitem {SME2}V. A. Kosteleck\'{y} and S. Samuel, Phys. Rev. Lett. 63, 224
(1989); Phys. Rev. D 40, 1886 (1989); 39, 683 (1989); Phys. Rev. Lett. 66,
1811 (1991); V. A. Kosteleck\'{y} and R. Potting, Phys. Rev. D 51, 3923 (1995).

\bibitem {Bazeia1}M. N. Barreto, D. Bazeia, and R. Menezes, Phys. Rev. D 73,
065015 (2006).

\bibitem {Dutra1}A. de Souza Dutra, M. Hott, and F. A. Barone, Phys. Rev. D
74, 085030(2006).

\bibitem {Gomes1}D. Bazeia, {M. M. Ferreira. Jr.}, A. Gomes, and R. Menezes, Physica D:
Nonlinear Phenomena 239, 942 (2010).

\bibitem {Dutra2}A. de Souza Dutra and R. A. C. Correa, Phys. Rev. D 83,
105007 (2011).

\bibitem {Seifert}M. D. Seifert, Phys. Rev. Lett. 105, 201601 (2010); Phys.
Rev. D 82, 125015 (2010).

\bibitem {Baeta}A. P. Ba\^{e}ta Scarpelli, H. Belich, J. L. Boldo, and J. A.
Helay\"{e}l-Neto, Phys. Rev. D 67, 085021 (2003).

\bibitem {Lazar}R. Casana, G. Lazar, Phys. Rev. D 90, 065007 (2014).

\bibitem {Miller1}C. Miller, R. Casana, M. M. Ferreira Jr., and E. da Hora,
Phys. Rev. D 86, 065011 (2012).

\bibitem {Miller2}R. Casana, M. M. Ferreira Jr., E. da Hora, and C. Miller,
Phys. Lett. B 718, 620 (2012).

\bibitem {Hott}C.H. Coronado Villalobos, J.M. Hoff da Silva, M.B. Hott, H.
Belich, Eur. Phys. J. C 74, 27991 (2014).

\bibitem {Belich}H. Belich, F.J.L. Leal, H.L.C. Louzada, M.T.D. Orlando,
Phys.Rev. D 86, 125037 (2012).

\bibitem {Sourrou1}L. Sourrouille, Phys. Rev. D 89, 087702 (2014).

\bibitem {Sourrou2}R. Casana and L. Sourrouille, Phys. Lett. B 726, 488 (2013).

\bibitem {Alyson}R. Casana, M. M. Ferreira Jr., E. da Hora, A. B. F. Neves,
Eur. Phys. J. C74, 3064 (2014).

\bibitem {Correa1}A. de Souza Dutra, R. A. C. Correa, Adv. High Energy Phys.
2015, 673716 (2015).

\bibitem {Correa2}R. A. C. Correa, R. da Rocha, A. de Souza Dutra, Ann. Phys.
359, 198~(2015).

\bibitem {Correa3}R. A. C. Correa, Roldao da Rocha, A. de Souza Dutra, Phys.
Rev. D 91, 125021 (2015).

\bibitem {Torres}M. Torres, Phys. Rev. D 46, 2295 (1992); J. Escalona, M.
Torres, and A. Antill\'{o}n, Mod. Phys. Lett. A 08, 2955 (1993).

\bibitem {Ghosh}P. K. Ghosh, Phys. Rev. D 49, 5458 (1994).

\bibitem {LM}T. Lee and H. Min, Phys. Rev. D 50, 7738 (1994); M. Torres, Rev.
D 51, 4533 (1995); A. Antill\'{o}n, J. Escalona, and M. Torres, Phys. Rev. D
55, 6327 (1997); F. Chandelier, Y. Georgelin, M. Lassaut, T. Masson, and
J.C.Wallet, Phys. Rev. D 70, 065016 (2004).

\bibitem {petrov1}L. H. C. Borges, A. G. Dias, A. F. Ferrari, J. R.
Nascimento, A. Yu. Petrov, Phys. Rev. D 89, 045005 (2014).

\bibitem {petrov2}L. H. C. Borges, A. G. Dias, A. F. Ferrari, J. R.
Nascimento, A. Yu. Petrov, Phys. Lett. B 756, 332 (2016).

\bibitem {KMNM1}V.A. Kostelecky and M. Mewes, Phys. Rev. 80, 015020 (2009); M.
Schreck, Phys. Rev. 89, 105019 (2014); M. Cambiaso, R. Lehnert, R. Potting,
Phys. Rev. D \textbf{85}, 085023 (2012); M. Schreck, Phys. Rev. D \textbf{89,}
105019 (2014) ; Phys. Rev. D \textbf{90}, 085025 (2014); B. Agostini, F. A.
Barone, F. E. Barone, P. Gaete, J. A. Helay\"{e}l-Neto, Phys. Lett. B
\textbf{708}, 212 (2012); L. Campanelli, Phys. Rev. D \textbf{90}, 105014
(2014); R. Bufalo, B.M. Pimentel, D.E. Soto, Physical Review D \textbf{90},
085012 (2014).

\bibitem {KMNM2}V.A. Kostelecky and M. Mewes, Phys. Rev. 88, 096006 (2013); M.
Schreck, Phys. Rev. 90, 085025 (2014).

\bibitem {NM1}H. Belich, T. Costa-Soares, M. M. Ferreira, Jr., and J. A.
Helay\"{e}l-Neto, Eur. Phys. J. C 41, 421 (2005).

\bibitem {Charneski}B. Charneski, M. Gomes, R.V. Maluf, and A. J. da Silva,
Phys. Rev. D 86, 045003 (2012).

\bibitem {Petrov}T. Mariz, J. R. Nascimento, A.Y. Petrov, Phys. Rev. D 85,
125003 (2012); G. Gazzola, H. G. Fargnoli, A. P. Ba\^{e}ta Scarpelli, Marcos
Sampaio, M. C. Nemes, J. Phys. G 39, 035002 (2012); A. P. Baeta Scarpelli, J.
Phys. G \textbf{39}, 125001 (2012); L. C. T. Brito, H. G. Fargnoli, and A. P.
Baeta Scarpelli, Phys. Rev. D \textbf{87}, 125023 (2013).

\bibitem {Bakke}K. Bakke and H. Belich, J. Phys. G 39, 085001 (2012); K.Bakke,
H. Belich, and E. O. Silva, J. Math. Phys. (N.Y.) 52, 063505 (2011); J. Phys.
G 39, 055004 (2012); Ann. Phys. (Berlin) 523, 910 (2011); K. Bakke and H.
Belich, Eur. Phys. J. Plus 127, 102 (2012).

\bibitem {NMACB}H. Belich, E. O. Silva, M. M. Ferreira, Jr., and M. T. D.
Orlando, Phys. Rev. D 83, 125025 (2011).

\bibitem {claudio1}R. Casana, C. F. Farias, and M. M. Ferreira, Jr., Phys.
Rev. D \textbf{92}, 125024 (2015).

\bibitem{claudio2}R. Casana, C. F. Farias, M. M. Ferreira Jr., and G. Lazar, {Phys. Rev. D \textbf{94}, 065036 (2016).}
\end{thebibliography}
\end{document}